# The evolution of Zipf's law indicative of city development

Yanguang Chen

(Department of Geography, College of Urban and Environmental Sciences, Peking University, Beijing 100871, P.R. China. E-mail: chenyg@pku.edu.cn)

**Abstract**: Zipf's law of city-size distributions can be expressed by three types of mathematical models: one-parameter form, two-parameter form, and three-parameter form. The one-parameter and one of the two-parameter models are familiar to urban scientists. However, the three-parameter model and another type of two-parameter model have not attracted attention. This paper is devoted to exploring the conditions and scopes of application of this Zipf models. By mathematical reasoning and empirical analysis, new discoveries are made as follows. First, if the size distribution of cities in a geographical region cannot be described with the one- or two-parameter model, maybe it can be characterized by the three-parameter model with a scaling factor and a scale-translational factor. Second, all these Zipf models can be unified by hierarchical scaling laws based on cascade structure. Third, the patterns of city-size distributions seems to evolve from three-parameter mode to two-parameter mode, and then to one-parameter mode. Four-year census data of China's cities are employed to verify the three-parameter Zipf's law and the corresponding hierarchical structure of rank-size distributions. This study is revealing for people to understand the scientific laws of social systems and the property of urban development.

# 1. Introduction

The evolutional process of the mathematical expressions of Zipf's law is from one parameter to two parameters, and then to three parameters. The original form of Zipf's law is presented by Auerbach (1913), and it is a one-parameter model (Carroll, 1982; Zipf, 1949). Ten years later, a two-parameter function appeared (Aitchison and Brown, 1957; Lotka, 1925; Singer, 1936), and it evolved into the common form of Zipf's law. A more general expression is what is called three-parameter Zipf's model (Chen and Zhou, 2003; Gabaix, 1999a; Gell-Mann, 1994). Mandelbrot (1982) derived a three-parameter Zipf model using the ideas from fractal, Winiwarter (1983) suggested a practical form of the three-parameter expression, Chen (2001) derived a three-parameter form from a pair of exponential laws of hierarchies of cities, and Gabaix and Ibragimov (2011) made a suggestion that the third parameter should equal 1/2. Today, the one-parameter model and one of the two-parameter Zipf's models have been widely used to make urban and economic studies (Carroll, 1982; Batty and Longley, 1994; Chen and Zhou, 2004; Frankhauser, 1998; Gabaix, 1999b; Gabaix and Ioannides, 2004; Jiang and Jia, 2011; Jiang and Yao, 2010; Krugman, 1996; Zhou, 1999). However, the spheres of application and physical meaning of the three-parameter Zipf model and the corresponding two-parameter model are not yet clear.

In fact, there are varied models that can be employed to describe the rank-size distribution of cities. Zipf's law is the most probable one. Whether or not Zipf's model is a universal law for cities, urban scientists have different viewpoints (Jiang *et al*, 2015). I agree with Gabaix and Ioannides (2004) who once said: "The main question of empirical work should be how well a theory fits, rather than whether or not it fit perfectly (i.e. within the standard errors)." A discovery is that the rank-size distributions of cities in different geographical regions (e.g., countries, states, provinces) can be characterized by different Zipf's models. If the one-parameter Zipf model cannot be well fitted to a set of observational data of city sizes, the two-parameter Zipf models will do well; and if the two-parameter model cannot be well fitted to the observational dataset of an urban system, the three-parameter Zipf model will fit it well. The rank-size distribution is actually of evolutional process rather than a determined pattern. The models of Zipf's law may be associated with stages of urbanization, and at different stages, we need different models.

Urban geography is different from classical physics, and the urban laws are always of scale-translational or scaling symmetry rather than spatio-temporal translational symmetry. At different places or different time, the mathematical forms of a geographical law may be of subtle distinction. This paper is devoted to exploring the similarities and differences between variable

Zipf models. The rest of the article is arranged as follows. In section 2, three types of Zipf's models are compared with one another, and the relationships between the three-parameter Zipf model and the hierarchical scaling law are demonstrated by mathematical transform. In section 3, four-year census data of cities are employed to validate the three-parameter Zipf model and the corresponding hierarchical scaling law. In section 4, several related questions are discussed, and finally, the paper is concluded by summarizing the main points of this study. The novelty of this study is as follows. First, the mathematical structure of the three-parameter Zipf's law is made clear. Second, the corresponding relationships between different Zipf models and different self-similar hierarchies are clarified. Third, a hypothesis about the evolutional process of different Zipf distributions is proposed for understanding urbanization.

## 2. Models

### 2.1 Three Zipf model of city-size distributions

The developing course of Zipf's law is as follows: from one-parameter mode to two-parameter mode, and then to three-parameter model. On the contrary, the evolution of city-size distribution seem to be as below: from three-parameter pattern to two-parameter pattern, and then to one-parameter pattern. Zipf's law was originally expressed as a one-parameter model, that is

$$P(r) = \frac{P_1}{r}, \tag{1}$$

where $r$ refers to city rank ($r$=1, 2, 3, …), $P(r)$ to the size of the $r$th city, and the only parameter is the proportionality coefficient $P_1$, denoting the size of the largest city. Equation (1) represents the pure form of Zipf's law, which was afterwards generalized to the following two-parameter expression

$$P(r) = P_1 r^{-q}, \tag{2}$$

in which $q$ denotes a scaling exponent, the other symbols are the same as in equation (1). There are two parameters in equation (2): one is the proportionality coefficient, and the other is the scaling exponent. The two-parameter Zipf model is more practical than the one-parameter Zipf model, but the one-parameter model is more famous than the two-parameter expression.

The three-parameter Zipf's law was earlier proposed by Mandelbrot (1982), who derived the model by using the idea from fractals. Winiwarter (1983) found a similar rank-size rule. The three-parameter function can be expressed as below:

$$P(r) = C(r+\zeta)^{-q}, \tag{3}$$

where $\zeta$ is an adjusting parameter, and $C$ is a proportionality coefficient. The other symbols are the same as in equation (2). This model can be derived from the cascade structure of urban hierarchies (Chen and Zhou, 2003). I will demonstrate that the three-parameter can be re-expressed as

$$P(r) = P_{1-k}(r+k)^{-q}, \tag{4}$$

in which $k$ is a scale-translational parameter of city rank, and the parameter $P_{1-k}$ indicates the size of the $(1-k)$th city that is defined in the possible world rather than the real world. Equation (4) suggests an absence of the leading cities in the top levels of a hierarchy of cities. The parameter $k$ can be termed translational factor, while the exponent $q$ denotes a scaling factor. If the scaling exponent $q=1$, we will have another type of two-parameter Zipf model, $P(r)=C/(r+k)$ (Table 1).

Table 1 A simple comparison between four forms of Zipf's models

| Parameter number | Type | Expression | Proportionality Coefficient | Scaling exponent | Adjusting parameter |
|---|---|---|---|---|---|
| One | Pure form | $P(r)=P_1/r$ | $P_1$ | 1 | 0 |
| Two (I) | Common form | $P(r)=P_1/r^q$ | $P_1$ | $q$ | 0 |
| Two (II) | Special form | $P(r)=C/(r+k)$ | $P_{1-k}$ | 1 | $k$ |
| Three | General form | $P(r)=C/(r+k)^q$ | $P_{1-k}$ | $q$ | $k$ |

**2.2 Three-parameter Zipf's model and fractal hierarchy**

It can be proved that Zipf's law can be transformed into the hierarchical scaling law (Chen, 2012a). However, different types of Zipf models correspond to different hierarchies with cascade structure (Figure 1; Table 2). A hierarchical scaling law can be decomposed into two exponential laws: one is the city *number law*, and the other, the city *size law* (Chen, 2012b). The three-parameter Zipf model, equation (3) or (4), can be derived from two exponential laws of the cascade structure of hierarchies of cities. Suppose that there is an urban hierarchy consisting of $M$ levels of cities. The levels are numbered $m=1, 2, \cdots, M$. The cascade structure of a hierarchy of cities can be described with a pair of exponential functions as follows

$$f_m = f_1 \delta^{m-1}, \tag{5}$$

$$P_m = P_1 \lambda^{1-m}, \tag{6}$$

where $f_m$ refers to the number of cities in the $m$th level, $P_m$ to the average size of the $f_m$ cities. As for the parameters, $f_1$ denotes the number of the cities in the top level, $P_1$ indicates the average size of the top-level cities, $\delta = f_{m+1}/f_m$ refers to the number ratio ($\delta > 1$), and $\lambda = P_m/P_{m+1}$ to the size ratio ($\lambda > 1$). From equations (5) and (6) it follows a three-parameter Zipf model. Let $m \to x \in [0, \infty)$, then we have $dN(x)/dx \propto f_m$, where $N(x)$ denotes the cumulative number of different levels of cities. Thus equations (5) and (6) can be re-expressed as

$$\frac{dN(x)}{dx} \propto f(x) = f_1 \delta^{x-1}, \tag{7}$$

$$P(x) = P_1 \lambda^{1-x}, \tag{8}$$

Differentiating equation (8) with respect to $x$ yields

$$\frac{dP(x)}{dx} = -P_1 \lambda^{1-x} \ln \lambda. \tag{9}$$

Using equation (9) to divide equation (7) yields

$$\frac{dN(x)}{dP(x)} = \frac{f_1}{P_1 \ln \lambda} (\lambda \delta)^{x-1}. \tag{10}$$

Taking logarithm to the base $\lambda \delta$ on both sides of equation (8) gives

$$x - 1 = (\log_{\lambda\delta} P_1 - \log_{\lambda\delta} P(x))/\log_{\lambda\delta} \lambda = \log_{\lambda\delta} [\frac{P_1}{P(x)}]^{\ln(\lambda\delta)/\ln\lambda}, \tag{11}$$

Substituting equation (11) into equation (10) to eliminate $x-1$ yields

$$\frac{dN(x)}{dP(x)} = \frac{f_1}{P_1 \ln \lambda} [\frac{P_1}{P(x)}]^{1+\ln\delta/\ln\lambda} = \frac{f_1 P_1^{\ln\delta/\ln\lambda}}{\ln\lambda} P(x)^{-(1+\ln\delta/\ln\lambda)}. \tag{12}$$

Integrating $dN(x)$ over $P$ (from $P_1$ to $P$) gives

$$N(P) = \frac{f_1 P_1^{\ln\delta/\ln\lambda}}{\ln\lambda} \int_{P_1}^{P} P^{-(1+\ln\delta/\ln\lambda)} dP = \frac{f_1 P_1^{\ln\lambda/\ln\delta}}{\ln\delta}(P^{-\ln\delta/\ln\lambda} - P_1^{-\ln\delta/\ln\lambda}). \tag{13}$$

Let $N(P) = r$ represent rank, and $P = P(r)$ represent the size of the $r$th city. Rearrange equation (13) yields

$$P(r) = P_1(\frac{f_1}{\ln\delta})^{\ln\lambda/\ln\delta}(r + \frac{f_1}{\ln\delta})^{-\ln\lambda/\ln\delta}, \tag{14}$$

in which the parameters can be expressed as below:

$$C = P_1(\frac{f_1}{\ln\delta})^{\ln\lambda/\ln\delta}, \quad \zeta = \frac{f_1}{\ln\delta}, \quad q = \frac{\ln\lambda}{\ln\delta}. \tag{15}$$

Thus equation (14) changes to a three-parameter Zipf model

$$P(r) = C(r+\zeta)^{-q}, \qquad (16)$$

where $\zeta$ is an adjustable parameter. If the number ratio $\delta$ is small enough, according to the approximate formula based on the Taylor's series, we have $\ln\delta=\delta-1$. If $\delta=2$ as given, then $\zeta \approx f_1$, which will be verified by observational data.

**Table 2 The relationships between four types of Zipf's models and hierarchical structure**

| | Zipf's law | | Hierarchical scaling law | | |
|---|---|---|---|---|---|
| **Parameter number** | Type | Expression | Top level ($m=1$) | Total size $S_m$ | Scaling exponent ($q$) |
| One | Pure form | $P(r)=P_1/r$ | Present | Constant | $q=1$ |
| Two (I) | Common form | $P(r)=P_1/r^q$ | Present | Exponential | $q \neq 1$ |
| Two (II) | Special form | $P(r)=C/(r+k)$ | Absent | Constant | $q=1$ |
| Three | General form | $P(r)=C/(r+k)^q$ | Absent | Exponential | $q \neq 1$ |

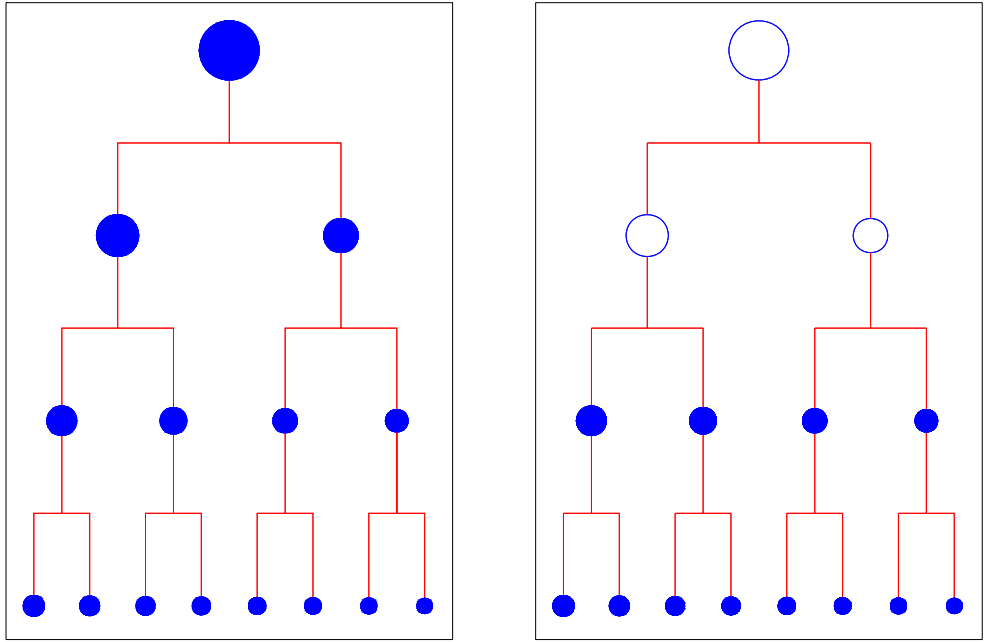

a. A hierarchy with top level  b. A hierarchy without top level

**Figure 1 Two schematic diagrams of urban hierarchies with cascade structure (the first four levels)** [**Note:** (1) The disks represent presence of cities, while the circles denote absence of cities. (2) Figure a corresponds to the one-parameter Zipf model and the common two-parameter Zipf model; Figure b corresponds to the three-parameter Zipf model and the special two-parameter Zipf model. ]

The three-parameter Zipf law is a fractal model of the rank-size distribution of cities. In fact, a hierarchical scaling law can be derived from equations (5) and (6), that is

$$f_m = \mu P_m^{-D}, \tag{17}$$

where $\mu = f_1 P_1^D$ denotes a proportionality coefficient, and $D = \ln(f_{m+1}/f_m)/\ln(P_m/P_{m+1})$ refers to a scaling exponent. By analogy with the similarity dimension formula, we can prove that the scaling exponent is associated with the fractal dimension of urban hierarchies. Integrating equation (5) over $x$ (from 1 to $m$) yields

$$N(m) = \int_1^m f(x)dx = f_1 \int_1^m \delta^{x-1} dx = \frac{f_1}{\ln \delta}(\delta^{m-1} - 1). \tag{18}$$

This suggests a number ratio as below:

$$\frac{N_{m+1}}{N_m} = \frac{N(m+1)}{N(m)} = \frac{\delta^m - 1}{\delta^{m-1} - 1} = \delta + \frac{\delta - 1}{\delta^{m-1} - 1}. \tag{19}$$

Because $\delta > 1$, we have $N_{m+1}/N_m \to \delta$. The reciprocal of the city size ratio is $P_{m+1}/P_m = 1/\lambda$. Thus we have

$$D = -\lim_{m \to \infty} \frac{\ln(N_{m+1}/N_m)}{\ln(P_{m+1}/P_m)} = -\frac{\ln(f_{m+1}/f_m)}{\ln(P_{m+1}/P_m)} = \frac{\ln \delta}{\ln \lambda} = \frac{1}{q}, \tag{20}$$

which is identical in form to the expression of the similarity dimension of fractals. Equation (20) suggests an inherent relationship between Zipf's law, Pareto's law, and the hierarchical scaling law.

**2.3 Parameter estimation and algorithms**

The results of parameter estimation always depend on algorithms. According to the theoretical demonstration shown above, the three-parameter Zipf model is equivalent to a hierarchical scaling as follows

$$P_m = \eta f_m^{-q}, \tag{21}$$

where $\eta = f_1 P_1^q$ refers to a proportionality coefficient, and $q = \ln(P_m/P_{m+1})/\ln(f_{m+1}/f_m)$ to Zipf scaling exponent, which is the reciprocal of the fractal dimension of urban hierarchies. Equation (21) is actually the inverse function of equation (17). In theory, we have $q = 1/D$; empirically, by the ordinary least square (OLS) method, we have $q = R^2/D$, where $R$ is the Pearson correlation coefficient (Chen and Zhou, 2003). By the reduced major axis (RMA) (Zhang and Yu, 2010), we

have

$$D^* = \sqrt{D/q},\qquad(22)$$

where $D$ and $q$ are both the OLS-based estimated parameters, and $D^*$ denote the RMA-based estimated fractal dimension. The less significant the difference between $D$ and $D^*$, the more the hierarchies of cities follow the hierarchical scaling law.

## 3. Empirical evidence

### 3.1 Study area and methods

The three-parameter Zipf model can be employed to study the rank-size distribution of Chinese cities. Whether or not the size distribution of China's cities follows Zipf's law is a pending question, to which there are different answers. Some scholars say "yes" (Chen *et al*, 1993; Chen and Zhou, 2008; Gangopadhyay and Basu, 2009; Ye and Xie, 2012), while some scholars say "no" (Anderson and Ge, 2005; Benguigui and Blumenfeld-Lieberthal, 2007a; Benguigui and Blumenfeld-Lieberthal, 2007b). Four datasets of city sizes are available for this research, including the observations of the third census (1982), the fourth census (1990), the fifth census (2000), and the sixth census (2010). Among these datasets, the 2000-year one was processed and put to rights by Zhou and Yu (2004a; 2004b). The analytical process is as follows. **Step 1**, identify scaling ranges. Using a double logarithmic plot, we can determine a scale-free range for a rank-size distribution. A scaling range is a straight line segment on the log-log plot, which indicates a scale-free extent of the Zipf distribution. Beyond the scale-free range, the data points form a droopy tail representing undeveloped cities. **Step 2**, search the adjusting parameters. Based on the scaling range, we can make a least square calculation by means of the logarithmic values of rank and size. Changing the scale parameter $k$, we will have different values of goodness of fit ($R^2$). If the $R$ square reaches its maximum, the search process ceases. **Step 3**, build models. Given $k$, it follows a three-parameter Zipf model of cities. **Step 4**, validate models. A three-parameter Zipf model can be confirmed by the corresponding hierarchical structure and scaling relation.

### 3.2 Calculations

The three-parameter Zipf model can be fitted to the census data of Chinese city sizes as a whole.

However, the rank-size patterns seem to emerge from the evolutional process because the scaling ranges becomes longer and longer and the values of goodness of fit goes up and up over time (Table 3). There were 238 cities in 1982, but only about 150 cities are within the scaling range (Figure 2(a)). For this dataset, the searching calculation based on the scaling range fails to converge. Therefore we cannot find a valid adjusting parameter. Finally, an estimated parameter value is set as $k=4$, which corresponds to the scaling exponent $q=1.0515$. For the datasets of 1990, 2000, and 2010, the city numbers are 460, 666, and 654, and the city numbers within the scaling ranges are about 350, 550, and 550. The calculation of parameter searching converge, and we have $k=4$ for 1990, $k=5$, for 2000, and $k=4$ for 2010. The corresponding scaling exponent values are $q=0.9390$, $q=0.9702$, and $q=1.0056$ (Figure 2(b, c, d)).

Table 3 The parameters and statistics of the three-parameter Zipf's model of China's cities

| Item | 1982 | 1990 | 2000 | 2010 |
|---|---:|---:|---:|---:|
| City number | 238 | 460 | 666 | 654 |
| Scaling range | 1-150 | 1-350 | 1-550 | 1-550 |
| Translational factor $k$ | 4 | 4 | 5 | 4 |
| Coefficient $P_1$ | 34385461.9592 | 26191080.8940 | 55117768.7109 | 82392190.5267 |
| Scaling exponent $q$ | 1.0515 | 0.9390 | 0.9702 | 1.0056 |
| Standard error $s$ | 0.0109 | 0.0031 | 0.0020 | 0.0030 |
| Goodness of fit $R^2$ | 0.9842 | 0.9963 | 0.9977 | 0.9953 |
| Fractal dimension $D$ | 0.9360 | 1.0609 | 1.0283 | 0.9897 |

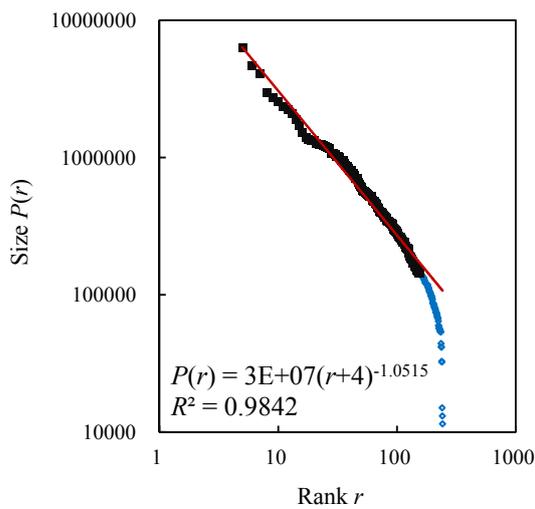

a. 1982

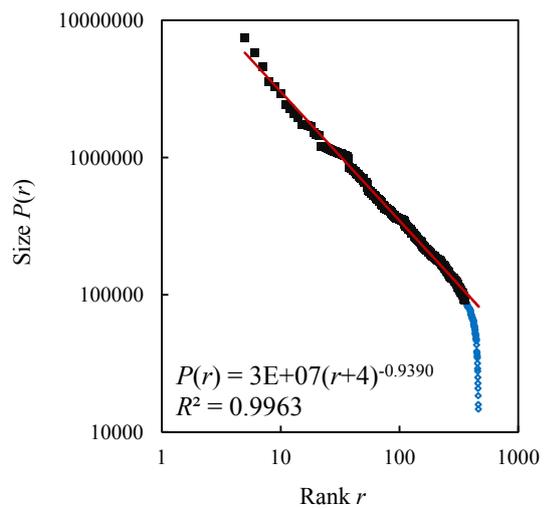

b. 1990

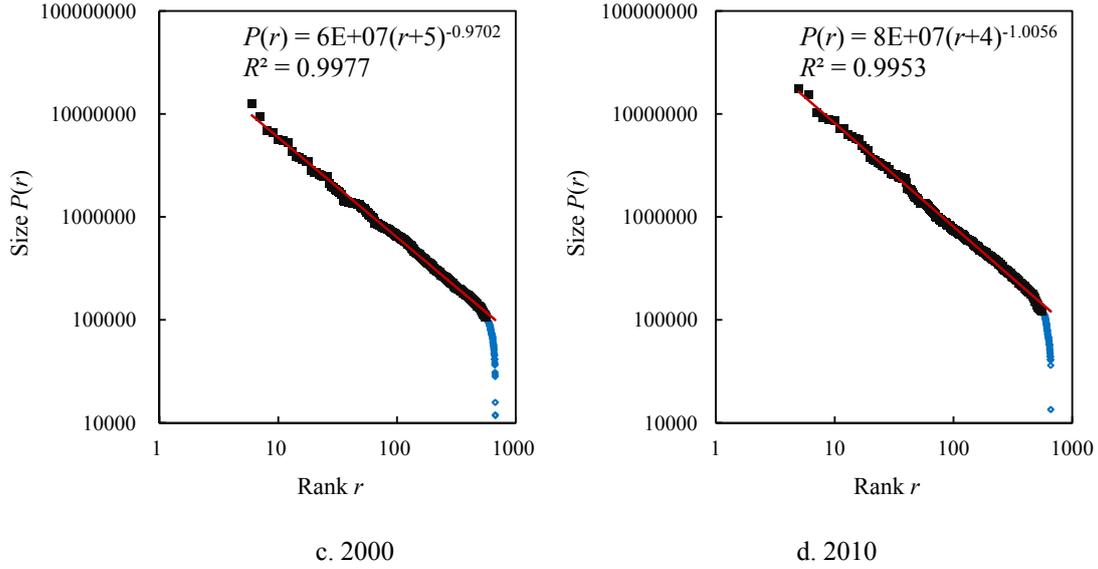

c. 2000　　　　　　　　　　　　　d. 2010

**Figure 2 The rank-size distributions of China's cities based on the three-parameter Zipf's model**

The three-parameter patterns of size distributions can be validated by the cascade structure of urban hierarchy. There are two approaches to constructing models of hierarchy of cities: one is based city size, and the other, based on city number (Chen and Zhou, 2003; Chen and Zhou, 2004; Jiang and Yao, 2010). The former is based on the $2^n$ rule of Davis (1978) and the model of city hierarchies proposed by Beckmann (1958) and Liang (1999), while the latter is the symmetric form of the former. The size-based hierarchy can be used to identify the primate distribution, while number-based hierarchy can be used to identify the scale-translational Zipf distribution. Given size ratio, e.g., $\lambda=2$, it follows that the size scale will be a geometric sequence such as $P_1, P_1/2, P_1/4, \ldots, P_1/2^{m-1},\ldots$. Then, the number of cities coming between $P_1/2^{m-1}$ and $P_1/2^m$ can be counted ($P_1/2^m \leq P_r < P_1/2^{m-1}$), and thus we have a size-based urban hierarchy. On the other hand, if number ratio, e.g., $\delta=2$, is given, then the city number of different levels will be a geometric sequence such as 1, 2, 4, …, $2^{m-1}$. The lower limit of city size or the average size at each class can be easily reckoned, and thus we have a number-based urban hierarchy. The dual models can be employed to describe the deep cascade structure of urban hierarchies. The bottom level of a hierarchical model is usually misshapen because of incomplete datasets or undergrown small cities and towns, no matter what types of approaches will be adopted to make the urban hierarchy. The uncompleted bottom level is termed lame-duck class (Davis, 1978). In a log-log plot, the lame-duck class takes on an outlier, showing an exceptional value.

The number-based hierarchy modeling can be employed to identify the three-parameter rank-size distribution. In 1982, 238 cities were officially counted in China's census dataset. The 238 cities can be divided into 6 levels, and the bottom level (114 cities) is a lame-duck class. At least, the first and second levels are absent (Table 4). In 1990, 460 cities were reckoned in Chinese official census dataset. The 460 cities can be grouped under 7 heads, and the last level (208 cites) is a lame-duck class. The topmost level and sublevel levels are of undergrowth (Table 5). In 2000, 666 cities were on record. The 666 cities can be classified into 8 levels, and the last level (158 cites) is a lame-duck class. There is an absence of the first two levels (Table 6). In 2010, some large cities and nearby small cities were merged to form outsize cities, and thus 654 cities were recorded in the official dataset. The 654 cities can be put into 8 levels, and the last level (146 cites) is still a lame-duck class. The first two levels are not yet developed (Table 7).

Table 4 The statistics and parameters of the 1982-year hierarchy of China's cities

| Level $m$ | City number $f_m$ | Total population $S_m$ | Average population $P_m$ | Size ratio $r_p$ |
|---|---|---|---|---|
| 1 | 1 | -- | -- | -- |
| 2 | 2 | -- | -- | -- |
| 3 | 4 | 18096531 | 4524132.7500 | -- |
| 4 | 8 | 17081306 | 2135163.2500 | 2.1189 |
| 5 | 16 | 19214036 | 1200877.2500 | 1.7780 |
| 6 | 32 | 21630428 | 675950.8750 | 1.7766 |
| 7 | 64 | 19619594 | 306556.1563 | 2.2050 |
| *8* | *114* | *12232124* | *107299.3333* | *2.8570* |

**Note**: The average total size is 19128379, the average size ratio is 1.9696, and the estimated scaling exponent $q$=ln(1.9696)/ln(2)=0.9779. The top and the second classes are absent, and the bottom class is incomplete.

Table 5 The statistics and parameters of the 1990-year hierarchy of China's cities

| Level $m$ | City number $f_m$ | Total population $S_m$ | Average population $P_m$ | Size ratio $r_p$ |
|---|---|---|---|---|
| 1 | 1 | -- | -- | -- |
| 2 | 2 | -- | -- | -- |
| 3 | 4 | 21417517 | 5354379.2500 | -- |
| 4 | 8 | 18431945 | 2303993.1250 | 2.3240 |
| 5 | 16 | 20330183 | 1270636.4375 | 1.8133 |
| 6 | 32 | 23088645 | 721520.1563 | 1.7611 |
| 7 | 64 | 23809117 | 372017.4531 | 1.9395 |
| 8 | 128 | 25535721 | 199497.8203 | 1.8648 |
| *9* | *208* | *18640561* | *89618.0817* | *2.2261* |

**Note**: The average total size is 22102188, the average size ratio is 1.9405, and the estimated scaling exponent

$q=\ln(1.9405)/\ln(2)=0.9564$. The top and the second classes are absent, and the bottom class is incomplete.

Table 6 The statistics and parameters of the 2000-year hierarchy of China's cities

| Level $m$ | City number $f_m$ | Total population $S_m$ | Average population $P_m$ | Size ratio $r_p$ |
|---|---|---|---|---|
| 1 | 1 | -- | -- | -- |
| 2 | 2 | -- | -- | -- |
| 3 | 4 | 35635578 | 8908894.5000 | -- |
| 4 | 8 | 35692935 | 4461616.8750 | 1.9968 |
| 5 | 16 | 37381398 | 2336337.3750 | 1.9097 |
| 6 | 32 | 39414749 | 1231710.9063 | 1.8968 |
| 7 | 64 | 43773412 | 683959.5625 | 1.8009 |
| 8 | 128 | 45049897 | 351952.3203 | 1.9433 |
| 9 | 256 | 45699688 | 178514.4063 | 1.9716 |
| 10 | 158 | 13673752 | 86542.7342 | 2.0627 |

**Note**: The average total size is 40378236.7143, the average size ratio is 1.9198, and the estimated scaling exponent $q=\ln(1.9198)/\ln(2)= 0.9410$. The top and the second classes are absent, and the bottom class is incomplete.

Table 7 The statistics and parameters of the 2010-year hierarchy of China's cities

| Level $m$ | City number $f_m$ | Total population $S_m$ | Average population $P_m$ | Size ratio $r_p$ |
|---|---|---|---|---|
| 1 | 1 | -- | -- | -- |
| 2 | 2 | -- | -- | -- |
| 3 | 4 | 52784771 | 13196192.7500 | -- |
| 4 | 8 | 55788148 | 6973518.5000 | 1.8923 |
| 5 | 16 | 54497801 | 3406112.5625 | 2.0474 |
| 6 | 32 | 54539028 | 1704344.6250 | 1.9985 |
| 7 | 64 | 51947700 | 811682.8125 | 2.0998 |
| 8 | 128 | 57120905 | 446257.0703 | 1.8189 |
| 9 | 256 | 56728600 | 221596.0938 | 2.0138 |
| 10 | 146 | 14453458 | 98996.2877 | 2.2384 |

**Note**: The average total size is 54772421.8571, the average size ratio is 1.9784, and the estimated scaling exponent $q=\ln(1.9784)/\ln(2)= 0.9844$. The top and the second classes are absent, and the bottom class is incomplete.

It can be demonstrated that the hierarchies of China's cities follow the hierarchical scaling law. The first two levels are absent, and the last level is undergrown. Therefore, the scaling range is from the third level to the penult level (the last but one class). If the last levels are eliminated as outliers, the hierarchical scaling laws, equation (17) or equation (21), can be well fitted to the hierarchical datasets, which are displayed in Tables 4, 5, 6, and 7. The values of goodness of fit are all greater than 0.996 (Table 8). All the data points but the exceptional value indicating the lame-duck class form a straight line approximately in a log-log plot (Figure 3). The Zipf scaling exponents indicated by the slopes are less than 1. Accordingly, the fractal dimension values are greater than 1 in a way.

Table 8 The parameters and statistics of the hierarchical scaling model of China's cities

| Item | 1982 | 1990 | 2000 | 2010 |
|---|---|---|---|---|
| **Level number** | 6 | 7 | 8 | 8 |
| **Scaling range** | 3-7 | 3-8 | 3-9 | 3-9 |
| **Coefficient $P_1$** | 16263149.8127 | 17493469.3827 | 31456936.9920 | 52676754.5261 |
| **Scaling exponent $q$** | 0.9426 | 0.9269 | 0.9294 | 0.9889 |
| **Standard error $s$** | 0.0331 | 0.0267 | 0.0077 | 0.0094 |
| **Goodness of fit $R^2$** | 0.9963 | 0.9967 | 0.9997 | 0.9996 |
| **Fractal dimension $D$** | 1.0570 | 1.0753 | 1.0756 | 1.0108 |
| **Fractal dimension $D^*$** | 1.0589 | 1.0771 | 1.0757 | 1.0110 |

Note: The fractal dimension $D$ and the Zipf scaling exponent $q$ are given by the OLS method, while the fractal dimension $D^*$ is obtain by RMA method.

### 3.3 Analysis

Analyzing the above calculations based on the census data of urban population, we can reveal the deep structure of hierarchy and the rank-size distribution properties of China's cities. The main viewpoints on China's systems of cities can be summarized as follows.

**First, the rank-size distribution of Chinese cities follows hierarchical scaling law.** The scaling law is equivalent to a three-parameter Zipf's law ($k>0$, $q\neq1$). If we use the one-parameter ($q=1$) or the common two-parameter Zipf model ($k=0$, $q\neq1$) to describe China's city-size distribution, the effect is not satisfying. Maybe just because of this, some scholars did not believe that Chinese cities conform to Zipf's law (Anderson and Ge, 2005; Benguigui and Blumenfeld-Lieberthal, 2007a; Benguigui and Blumenfeld-Lieberthal, 2007b). However, if we find proper translational parameter values, we can model the size distributions of China's cities with the three-parameter Zipf's law and the corresponding hierarchical scaling law.

**Second, the regularity of the rank-size patterns of Chinese cities becomes more and more significant over time.** In 1982, the property of size distribution is not clear because we cannot find the most suitable value of translational parameter for Zipf's distribution. From 1990 on, the translational parameter values are determinate. The Zipf exponent goes up and up to 1, and accordingly, the fractal dimension of urban hierarchies goes down and down to 1. The total

population sizes of different classes take on linear growth (1990), and then exponential growth (2000), and finally, zero growth (2010) over order (*m*). This implies that the Zipf model have been evolving from the three-parameter mode to the special two-parameter mode ($k>0$, $q=1$). This lend further support to the suggestions that the geographical laws are the laws of evolution rather the laws of existence (Chen, 2014).

**Third, the extremes of the hierarchy of China's cities are absent or incomplete.** One the one hand, the top two levels is vacant because of absence of leading cities with powerful function. On the other hand, the bottom level is of undergrowth because that the small cities and towns did not receive due attention for a long time. China has a vast territory. It seems that no city's function is powerful enough to influence all the cities in China, despite the fact that Beijing and Shanghai have been regarded as what is called global cities by the Globalization and World Cities Research Network (GaWC) and the Chicago-based consulting firm A.T. Kearney and the Chicago Council on Global Affairs (http://en.wikipedia.org/wiki/Global_city). A global city, also called world city, is such a city that is generally thought of as an important node in the global economic network (Knox and Marston, 2009; Sassen, 1991; Taylor, 2001). China is a country of command economy rather than real market economy. The self-organization mechanism based on bottom-up evolution ("invisible hand") is restricted by the top-down action of government ("visible hand"). Under the circumstances, it is difficult for the small cities and towns to grow up. Therefore, there is drooping tail in each log-log plot of the rank-size distribution.

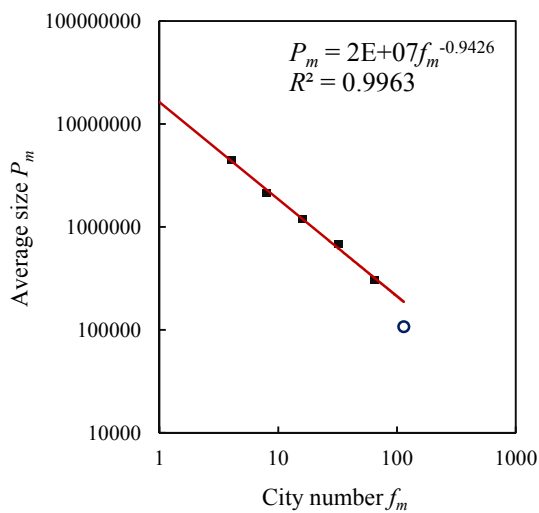
a. 1982

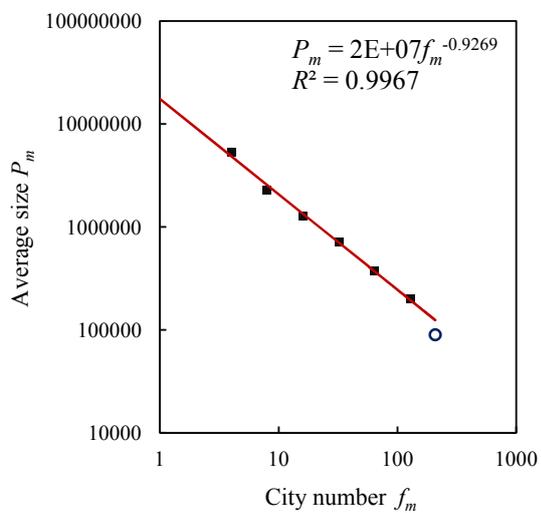
b. 1990

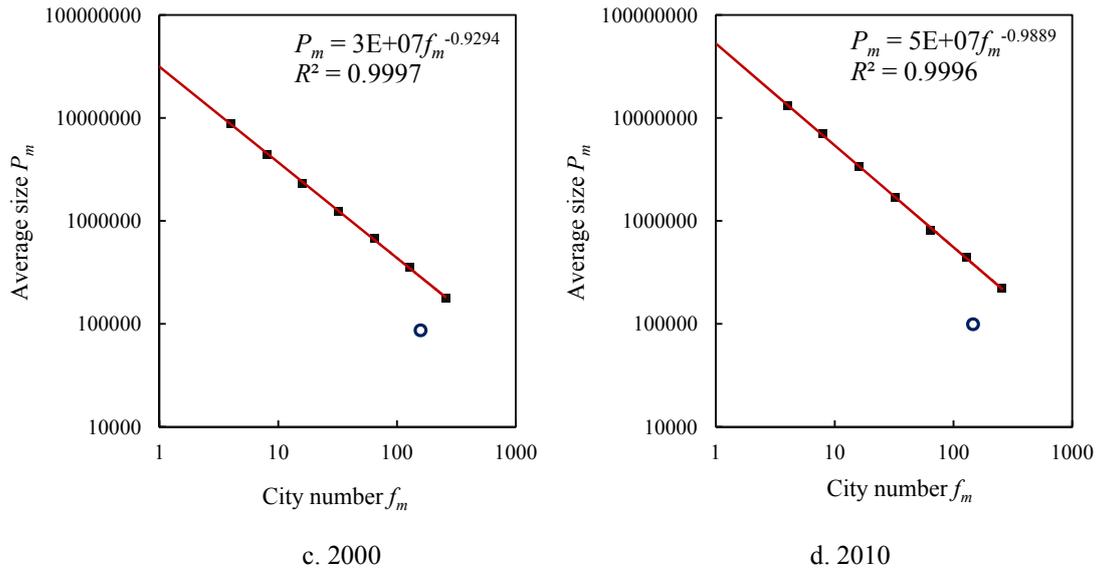

c. 2000   d. 2010

**Figure 3 The hierarchical scaling patterns of China's cities corresponding to the three-parameter Zipf's model** [**Note:** The small circles represent the lame-duck classes]

## 4. Discussion

The three-parameter model of Zipf's distribution of city sizes bears a good structure. It includes a proportion factor (constant coefficient), a scale factor (adjusting scale-translational parameter), and scaling factor (power exponent). The proportion factor indicates the size of the possible largest cities, the scale factor indicates a gap between the real largest city size and the possible largest city size, and the scaling factor indicates the consistency of a city with the largest city size. If a three-parameter Zipf distribution is converted into a hierarchy with cascade structure, the model and analytical process will be simplified because the scale-translational parameter will be eliminated and thus the three parameters will be reduced to two parameters (constant coefficient and scaling exponent). The inherent relationships between three types (four forms) of Zipf models and hierarchies with cascade structure are tabulated above (Table 2).

The rank-size distribution dominated by Zipf's law is a signature of hierarchical order of urban systems. Urban system is correlated with urbanization (Knox and Marston, 2009). The Zipf distribution of cities is a pattern emerged from the nonlinear dynamics of urbanization. Urbanization is a kind of phase transition, indicating a self-organizing process of regional population distribution from rural state to urban state. Fractals and Zipf's law are two indicators of self-organized criticality (SOC) (Bak, 1996; Batty and Xie, 1999; Chen and Zhou, 2008; Portugali, 2000). If the population transition evolves from an equilibrium state into a critical state by way of self-organization, fractal central places and rank-size patterns will emerge from the complex dynamics. By sigmoid curves, urbanization can be divided into four stages: *initial stage*, *acceleration stage*, *deceleration stage*, and *terminal stage*. The second and third stages can be merged into the *celerity stage*. Different stages of urbanization may coincide with different patterns of city rank-size distributions. Suppose there is a geographical region occupied by a system of cities and towns. The area of the region is $A_r$, and the area of the sphere of influence of the central city is $A_c$. If $A_c \approx A_r$, urbanization is in the self-organized critical state, and the rank-size distribution can be described by the one-parameter Zipf model. If $A_c \ll A_r$, urbanization is in the subcritical state, and the rank-size distribution can be described by the two- or three-parameter Zipf model. If $A_c \gg A_r$, urbanization is in the supercritical state, and the rank-size distribution will evolve into a primate distribution. In this case, the city-size distribution can be locally described by the two- or three-parameter Zipf model. If a rank-size distribution cannot be effectively described by any Zipf model, it indicates that urbanization is in a far from the critical state (Table 9). This implies that Zipf's law is actually an evolutional law, and the developing mechanism of rank-size distributions is involved with the complex dynamics of urbanization (Figure 4).

Table 9 The self-organizing states of urbanization and city-size distributions

| Urbanization | | City-size distributions | | |
|---|---|---|---|---|
| Stage | State | Distribution | Model | Scaling exponent |
| Pre-urbanization (initial stage) | Far from critical state | Other distribution, e.g., lognormal distribution | Non power law | No |
| Acceleration stage | Subcritical state | Zipf distribution | $P(r)=C/(r+k)^q$, $P(r)=C/(r+k)$ | $1/2<q<2$ |
| Deceleration stage | Critical state | Zipf distribution | $P(r)=P_1/r$ | $q=1$ |
| Post-urbanization (terminal stage) | Supercritical state | Primate distribution, local Zipf distribution | $P(r)=C/(r+k)$, $P(r)=P_1/r^q$ | $1/2<q<2$ |

As indicated above, Zipf's law is one of marks of SOC of complex systems (Bak, 1996). If the process of urbanization is near the self-organized critical state, the scaling exponent of the three-parameter Zipf model will be close to 1 ($k>1$, $q\rightarrow 1$) (Chen and Zhou, 2008). And thus, the corresponding hierarchical scaling law will be become as blow:

$$f_m P_m = f_1 P_1 = \mu, \tag{23}$$

where $\mu$ is a constant. If the process of urbanization is in the self-organized critical state, the scaling exponent of the two-parameter Zipf model will be close to 1 ($k=0$, $q\to1$), thus we have

$$f_m P_m = P_1, \tag{24}$$

where $P_1$ is a size of the largest city. Both equations (23) and (24) suggest a conservation law of urban hierarchies. Of course, this conservation law is also an evolutional law instead of an iron law. It may emerge from the critical state or the state between chaos and order. SOC and edge of chaos represents the different sides of the same coin of complex dynamics (Kauffman, 1993).

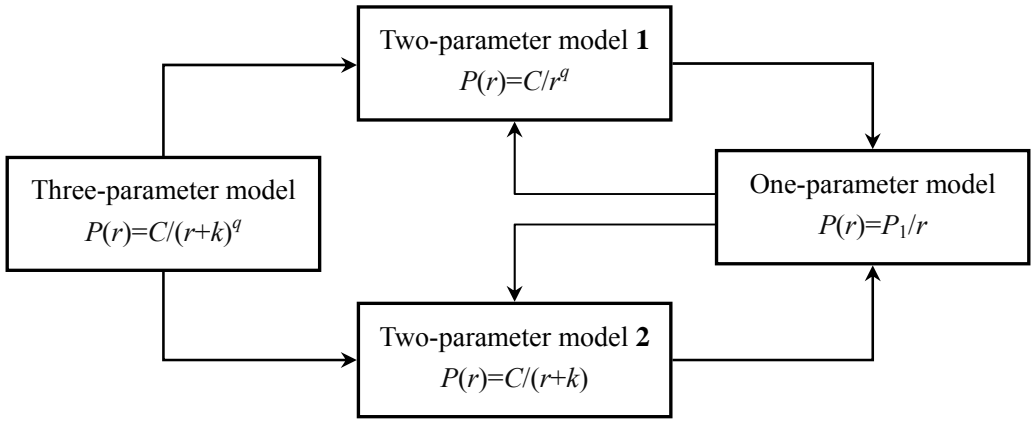

**Figure 4 A schematic diagram of the dynamic process and evolutional direction of Zipf's distributions**

The three-parameter Zipf law is appropriate for describing the size distribution of cities in large developing countries. In fact, it can be employed to model the rank-size distribution of Indian cities. By the 2011 census dataset, there is a record of 1207 cities. All the 1183 cities with population size greater than 20,000 fall into the scaling range. Fitting the three-parameter model to the data points within the scaling range yields

$$\hat{P}(r) = 75{,}962{,}650.6045(r+3)^{-1.1492}.$$

The goodness of fit is about $R^2=0.9969$, and the standard error of the Zipf exponent is about $s=0.0019$ (Figure 5(a)). These Indian cities can be organized into a hierarchy with cascade structure (Table 10). The hierarchical scaling relation is as below:

$$\hat{P}_m = 26,129,828.6091 f_m^{-1.0126}.$$

The goodness of fit is about $R^2$=0.9946, and the standard error of the scaling exponent is about $s$=0.0304 (Figure 5(b)).

Table 10 The statistics and parameters of the 2011-year hierarchy of India's cities

| Level $m$ | City number $f_m$ | Total population $S_m$ | Average population $P_m$ | Size ratio $r_p$ |
|---|---|---|---|---|
| 1 | 1 | -- | -- | -- |
| 2 | 2 | -- | -- | -- |
| 3 | 4 | 23486282 | 5871570.5000 | -- |
| 4 | 8 | 25487612 | 3185951.5000 | 1.8430 |
| 5 | 16 | 24145723 | 1509107.6875 | 2.1111 |
| 6 | 32 | 24554644 | 767332.6250 | 1.9667 |
| 7 | 64 | 30960225 | 483753.5156 | 1.5862 |
| 8 | 128 | 28114854 | 219647.2969 | 2.2024 |
| 9 | 256 | 23842053 | 93133.0195 | 2.3584 |
| 10 | 512 | 20109645 | 39276.6504 | 2.3712 |
| 11 | 185 | 3908694 | 21128.0757 | 1.8590 |

**Note**: If we do not consider the lame-duck class (bottom level), the average total size of different levels is 25087629.75, the average size ratio is 2.0627, and the estimated scaling exponent $q$=ln(2.0627)/ln(2)= 1.0445.

Urban physics coming between natural science and social science differs from classical physics. Social science involves three types of studies: behavioral study, canonical study, and axiological study (Krone, 1980). The behavioral study is also called positive study, and the canonical study is termed normative study (Behravesh, 2008). The former is to examine the real world by observational facts, while the latter is to explore the ideal world based on optimization ideas. As for the axiological study, it mainly provides theoretical criterions for evaluating the gap between status quo (real state) and objective (ideal state). The one-parameter Zipf model shows the optimized rank-size distribution of cities defined in ideal world. In contrast, the other Zipf models reflect the actual city-size distributions appearing in real world. The gap between the real distributions and the ideal distribution is just the problem that needs to be solved by specialists and managers of city planning and spatial optimization.

The chief shortcoming of this article rests with demographic census. Due to the scale-free property of urban form and city-size distribution (Batty and Longley, 1994; Frankhauser, 1994;

Mandelbrot, 1965), it is hard to find an urban boundary and a threshold of population size. Therefore, a city is in fact subjectively defined rather than objectively identified in many countries (Jiang and Yao, 2010; Zhou, 1995). The city of United States of America (USA) is statistically defined by urban experts, while the city of the People's Republic of China (PRC) is officially identified by China's central government. There are about 2,200 cities in Mainland China. However, only about 660 settlements are approved to be "cities" in recent years. Many Chinese cities were not taken into account in the previous census datasets. As a result, the rank-size patterns cannot reflect the completely real geographical information of China's city-size distributions. Despite this deficiency, the empirical analyses of this paper are reliable because Zipf's distribution is an open distribution. In particular, the case of Indian cities shows a circumstantial evidence for the rank-size distribution of China's cities.

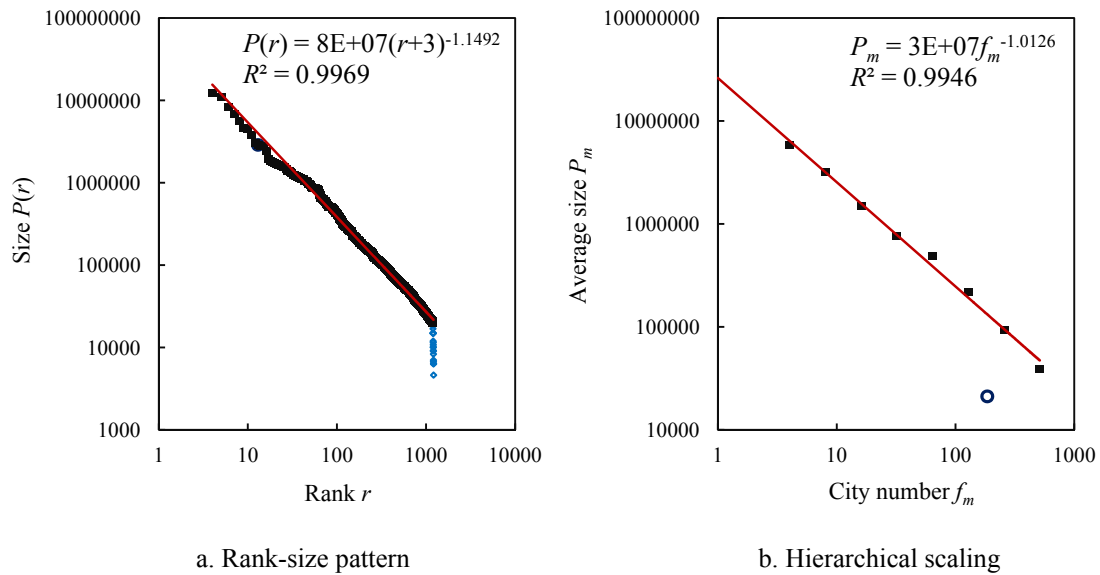

a. Rank-size pattern  b. Hierarchical scaling

**Figure 5 The three-parameter Zipf's distribution and the corresponding hierarchical scaling pattern of Indian cities (2011) [Note:** The data source is 2011 Census of India. In figure (a), the droopy tail represents the cities beyond the scaling range; in figure (b), the small circle represents the lame-duck classes**]**

## 5. Conclusions

Zipf's law of city-size distributions is a controversial question in urban studies. The mathematical models of Zipf distribution are developing models rather than fixed expression. The rank-size distributions are associated with the dynamic process of urbanization. In different stages of urbanization, maybe we need different models of Zipf's distribution. Based on the previous studies and discoveries made by many scholars, this paper proposes a more general model and an evolutional hypothesis for Zipf's law. From the theoretical derivation and empirical analyses, four main conclusions can be reached as follows.

**First, Zipf's law can be described with three types and four forms of models.** The first one is the one-parameter model with a proportion factor ($P_1$), the second one is common two-parameter model with a proportion factor ($P_1$) and a scaling factor ($q$), the third one is the special two-parameter model with a proportion factor ($C$) and a scale factor ($k$), and the fourth one is the three-parameter model with a proportion factor ($C$), a scale factor ($k$), and a scaling factor ($q$). The two-parameter Zipf models include two forms: one is the scale-dilation model ($q \neq 1$, $k=0$), and the other, scale-translational model ($q=1$, $k>0$). Among all these models, the most general one is the three-parameter model. The special two-parameter model can be treated as a generalized three-parameter model with a unit scaling exponent ($q=1$) in the studies of rank-size distributions.

**Second, all kinds of Zipf models can be transformed into the corresponding hierarchies with cascade structure.** For the one-parameter Zipf distribution, the hierarchy of cities has a top level ($k=0$), and the total urban population sizes in different levels approach a constant ($q=1$). For the common two-parameter Zipf distribution, the hierarchy has a top level ($k=0$), and the total population sizes in different levels approach to exponential increase ($q<1$) /decrease ($q>1$). For the special two-parameter Zipf distribution, the hierarchy has no top level ($k \geq 1$), and the total population sizes in different levels approach a constant ($q=1$). For the three-parameter Zipf distribution, the hierarchy has no top level ($k \geq 1$), and the total population sizes in different levels approach to exponential increase ($q<1$) /decrease ($q>1$). Because of the hierarchical structure, the rank-size distribution can be associated with self-organized networks of cities.

**Third, the Zipf's law and the hierarchical scaling law are valid within certain range of scales.** Many mathematical laws manifest a property of scale effect. If the scale is too large or too small, a mathematical law will be broken. If the largest city is not powerful enough to influence all the cities of the same geographical region, its size will be deviate downwards from the rank-size trend line. Thus the one-parameter Zipf model or the common two-parameter model cannot be well fitted to the observational data of city sizes. If the largest city is so powerful that its sphere of influence is larger than the geographical area of urban systems, the rank-size distribution will become a primate distribution. Thus the top level will go beyond the scaling range. On the other hand, because of undergrowth of small cities and towns, the bottom level will go beyond the scaling range.

**Fourth, different models of Zipf's law suggest a possible evolutional process of rank-size distributions of cities.** A hypothesis is that city size distributions evolve from the three-parameter mode to two-parameter mode, and finally, to one-parameter mode. Different parameters represent different conditions of city development. Zipf's distribution is a pattern emerging from the self-organizing process and appearing at the self-organized critical state. In short, Zipf's law is a law of evolution rather the laws of existence. The one-parameter Zipf model indicates the optimum size distribution of cities and suggests the optimized structure of urban hierarchies. If and only if urbanization evolve into a self-organized critical state, the Zipf distribution will emerge; and if and only if the rank-size distribution of cities evolve into the state of total system optimization, the pattern of the one-parameter Zipf distribution will come into being.

## Acknowledgements

This research was sponsored by the National Natural Science Foundation of China (Grant No. 41171129). The support is gratefully acknowledged. I would like to thank Dr. Jiejing Wang for data processing of China's city population.